\documentclass[showkeys,%
 preprint, 
 amsmath,amssymb,
 aps, physrev,
]{revtex4-2}

\usepackage{graphicx}
\usepackage{dcolumn}
\usepackage{bm}
\usepackage{soul}
\usepackage{xcolor} 
\sethlcolor{yellow} 
\usepackage{ulem}
\begin{document}


\title{\textbf{Evaluation of Coincidence Time Resolution in a liquid xenon detector with silicon photomultipliers.} 
}%

\author{N. Salor-Igui\~niz}
 \email{Corresponding author: nerea.salor@dipc.org}
\affiliation{Donostia International Physics Center, BERC Basque Excellence Research Centre, Manuel de Lardizabal 4, Donostia-San Sebasti\'an, E-20018, Spain}
\affiliation{Instituto de F\'isica Corpuscular (IFIC), CSIC \& Universitat de Val\`encia, Calle Catedr\'atico Jos\'e Beltr\'an, 2, Paterna, E-46980, Spain}

\author{J.M.~Benlloch-Rodr\'{i}guez}
\affiliation{Donostia International Physics Center, BERC Basque Excellence Research Centre, Manuel de Lardizabal 4, Donostia-San Sebasti\'an, E-20018, Spain}
\author{R.~Esteve}
\affiliation{Instituto de Instrumentaci\'on para Imagen Molecular (I3M), Centro Mixto CSIC - Universitat Polit\`ecnica de Val\`encia, Camino de Vera s/n, Valencia, E-46022, Spain}
\author{C.~Romo-Luque }
 \altaffiliation{Now at Los Alamos National Laboratory, US.}
\affiliation{Instituto de F\'isica Corpuscular (IFIC), CSIC \& Universitat de Val\`encia, Calle Catedr\'atico Jos\'e Beltr\'an, 2, Paterna, E-46980, Spain}
\author{R.J.~Aliaga}
\affiliation{ Instituto Universitario de Matem\'atica Pura y Aplicada (IUMPA), Universitat Polit\`ecnica de Val\`encia, Camino de Vera s/n, Valencia, E-46022, Spain}
\author{V.~\'Alvarez}
\affiliation{Instituto de Instrumentaci\'on para Imagen Molecular (I3M), Centro Mixto CSIC - Universitat Polit\`ecnica de Val\`encia, Camino de Vera s/n, Valencia, E-46022, Spain}
\author{F.~Ballester}
\affiliation{Instituto de Instrumentaci\'on para Imagen Molecular (I3M), Centro Mixto CSIC - Universitat Polit\`ecnica de Val\`encia, Camino de Vera s/n, Valencia, E-46022, Spain}
\author{R.~Gadea}
\affiliation{Instituto de Instrumentaci\'on para Imagen Molecular (I3M), Centro Mixto CSIC - Universitat Polit\`ecnica de Val\`encia, Camino de Vera s/n, Valencia, E-46022, Spain}
\author{F.~Monrabal}
\affiliation{Donostia International Physics Center, BERC Basque Excellence Research Centre, Manuel de Lardizabal 4, Donostia-San Sebasti\'an, E-20018, Spain}
\affiliation{Ikerbasque (Basque Foundation for Science), Bilbao, E-48009, Spain}
\author{M.~Querol}
\affiliation{Instituto de F\'isica Corpuscular (IFIC), CSIC \& Universitat de Val\`encia, Calle Catedr\'atico Jos\'e Beltr\'an, 2, Paterna, E-46980, Spain}
\author{M. Rappaport}
\affiliation{Weizmann Institute of Science, Herzl St 234, Rehovot, Israel}
\author{J.~Rodr\'iguez}
\affiliation{Instituto de Instrumentaci\'on para Imagen Molecular (I3M), Centro Mixto CSIC - Universitat Polit\`ecnica de Val\`encia, Camino de Vera s/n, Valencia, E-46022, Spain}
\author{J.~Rodr\'iguez-Ponce}
\affiliation{Instituto de F\'isica Corpuscular (IFIC), CSIC \& Universitat de Val\`encia, Calle Catedr\'atico Jos\'e Beltr\'an, 2, Paterna, E-46980, Spain}
\author{S.~Teruel-Pardo}
\affiliation{Instituto de F\'isica Corpuscular (IFIC), CSIC \& Universitat de Val\`encia, Calle Catedr\'atico Jos\'e Beltr\'an, 2, Paterna, E-46980, Spain}
\author{J.F.~Toledo}
\affiliation{Instituto de Instrumentaci\'on para Imagen Molecular (I3M), Centro Mixto CSIC - Universitat Polit\`ecnica de Val\`encia, Camino de Vera s/n, Valencia, E-46022, Spain}
\author{R.~Torres-Curado}
\affiliation{Instituto de Instrumentaci\'on para Imagen Molecular (I3M), Centro Mixto CSIC - Universitat Polit\`ecnica de Val\`encia, Camino de Vera s/n, Valencia, E-46022, Spain}
\author{P.~Ferrario}
 \altaffiliation{Also at Instituto de Física Corpuscular (Spain)}
 \altaffiliation{On leave.}
\affiliation{Donostia International Physics Center, BERC Basque Excellence Research Centre, Manuel de Lardizabal 4, Donostia-San Sebasti\'an, E-20018, Spain}
\affiliation{Ikerbasque (Basque Foundation for Science), Bilbao, E-48009, Spain}
\author{V.~Herrero}
\affiliation{Instituto de Instrumentaci\'on para Imagen Molecular (I3M), Centro Mixto CSIC - Universitat Polit\`ecnica de Val\`encia, Camino de Vera s/n, Valencia, E-46022, Spain}
\author{J.J.~G\'omez-Cadenas}
\affiliation{Donostia International Physics Center, BERC Basque Excellence Research Centre, Manuel de Lardizabal 4, Donostia-San Sebasti\'an, E-20018, Spain}
\affiliation{Ikerbasque (Basque Foundation for Science), Bilbao, E-48009, Spain}

\collaboration{PETALO Collaboration}

\date{\today}

\begin{abstract}
This work explores the combination of liquid xenon as a scintillating medium and silicon photomultipliers as a readout in Positron Emission Tomography (PET) for enhanced Time-Of-Flight resolution. We present the results of our first prototype optimized to maximize light collection using high-PDE, VUV-sensitive sensors and to minimize time fluctuations. We report a coincidence time resolution of 281 $\pm$ 2 ps FWHM, obtained using a $^{22}$Na calibration source. This result is competitive with the current state-of-the-art PET scanners and represents a significant step forward in the development of liquid xenon as a viable alternative to conventional scintillators in PET technology.
\end{abstract}

\maketitle


\section{Introduction}

Positron Emission Tomography (PET) is widely used in both clinical and research applications across different fields, such as oncology, cardiology, or neurology due to its ability to measure the in vivo biodistribution of a radiotracer with high sensitivity, making it an invaluable tool for exploring dynamic biological processes. The radiotracer typically involves a glucose molecule that has been modified by substituting one of its atoms with a radioactive isotope, such as $^{18}$F, which emits positrons. This modified glucose is preferentially taken up by areas of the body with higher metabolic activity, such as the brain or tumors, where it accumulates. Once emitted, the positron annihilates with an electron in the patient’s body, resulting in the production of two 511 keV gamma rays that travel in opposite directions. When the patient is positioned within a ring of scintillators, these gamma rays can be detected in coincidence, allowing for the formation of a line connecting them, known as a line of response (LoR).

A key feature in PET scanners is the timing information they capture. In conventional PET scanners, each detected coincidence event defines a LoR along which the annihilation event occurred. However, without additional timing information, all points along the LoR are equally likely to have emitted the gamma rays, and the location of the actual emission event cannot be determined individually. Instead, PET imaging focuses on accumulating a large number of such events and using the intersection of many LoRs to statistically reconstruct the spatial distribution of the radiotracer throughout the body.

\begin{figure}[htbp]
\centering 
\includegraphics[width=0.8\textwidth,]{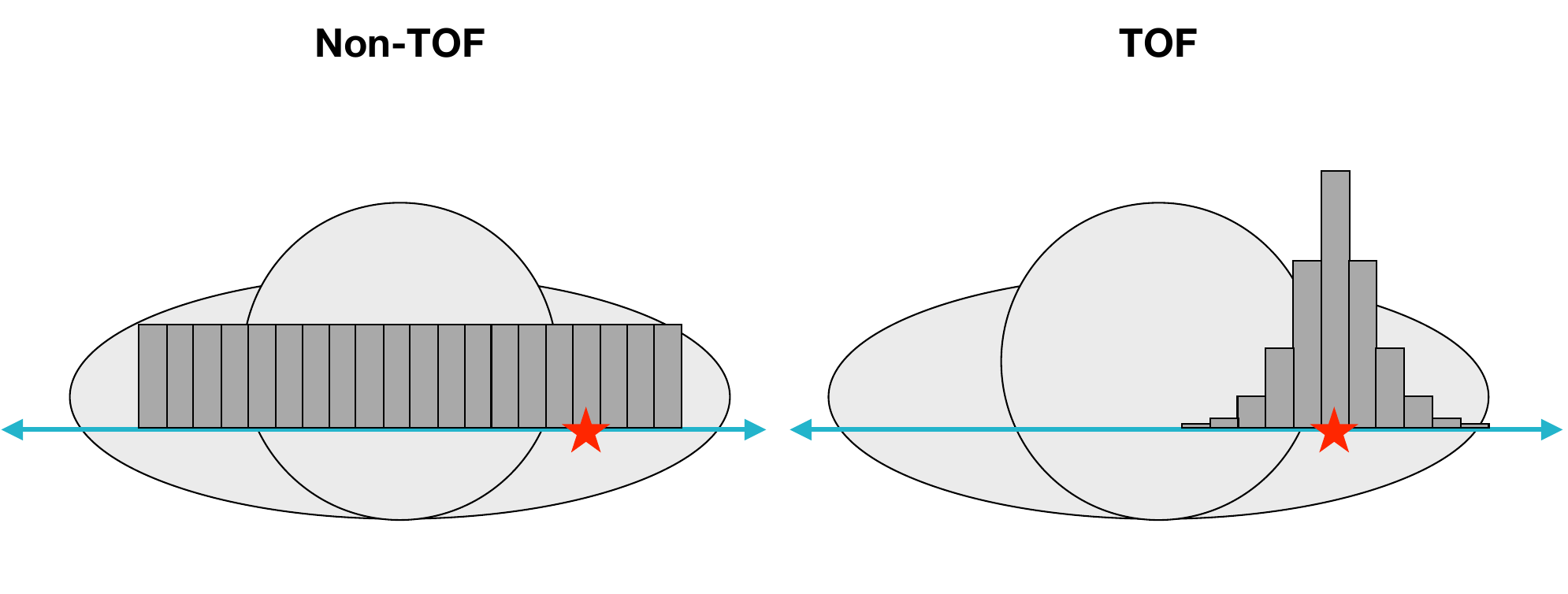}

\caption{\label{fig:resolutions} In a non-TOF reconstruction, all the segments in a single line of response contribute to the image with the same probability. In a TOF reconstruction, time resolution gives a location in the line of response with more probability of having emitted the gamma rays, allowing a better image resolution}.
\end{figure}

In contrast, PET scanners with Time-Of-Flight (TOF) capabilities enhance this process by measuring the time of arrival of the two gamma rays at the detectors with greater accuracy. This precise timing information is used to calculate the time difference between the two gamma rays, which correlates with the distance of the annihilation point from the center of the LoR. As a result, it is possible to identify a more localized segment along the LoR with a higher probability of emission (see Figure \ref{fig:resolutions}). This capability can improve image quality \cite{image-quality} or allow the same image quality to be achieved using lower radioisotope doses or shorter scan times.

The time resolution of PET scanners, known as coincidence time resolution or CTR, is determined primarily by the time resolution of the readout system and the scintillation decay time. These parameters must be as small as possible to maximize the number of events acquired per unit of time and to minimize the coincidence window used to correlate events on different detectors.

The concept of incorporating timing information into PET scanners originated in the late 1980s \cite{Mullani-1980}, when the first scanners were built \cite{Super-PETT, LETI, Texas, SP3, TTV03}. These early scanners used CsF and BaF$_2$ crystal scintillators, achieving time resolutions in the range of 450–750 ps FWHM \cite{CTR,CTR-2}. However, their low detection efficiency and limited light output resulted in poor spatial resolution and sensitivity, limiting their use mainly to research environments.
Over time, advancements in new scintillating materials, particularly LSO and LYSO, have significantly improved the TOF resolution. LSO and LYSO offer high light yields (30,000
photons/MeV and 25,000 photons/MeV, respectively) and fast decay times (both approximately 40 ns) \cite{LYSO,LSO}, enabling TOF resolutions in the range of 450-600 ps \cite{CTR-LYSO, CTR-LSO}. In particular, EXPLORER, the world’s first whole-body PET scanner, features a TOF resolution of 430 ps FWHM \cite{explorer}. More recently, the development of faster electronics and silicon photomultipliers (SiPMs) has further enhanced time resolution in TOF-PET, achieving values between 200 and 400 ps FWHM (see, for instance, References \cite{CTR-lowest, CTR-5ring}), depending on the characteristics of the crystals and SiPMs used.

The motivation behind improving the coincidence time resolution lies in its direct impact on the spatial localization of annihilation events along the LoR. Although the final image resolution depends on many reconstruction parameters, a rough spatial uncertainty can be estimated from the CTR using equation \ref{eq:position}, yielding values of ~64 mm for 430 ps and ~30 mm for 200 ps. This approximation helps illustrate how TOF information narrows the probable origin of the event even before full image reconstruction.

\begin{equation}
    \centering
    \Delta\text{x} = \frac{\text{c} \cdot \text{CTR} }{2} 
    \label{eq:position}
\end{equation}

Among the materials being explored to push time resolution even further, liquid xenon (LXe) stands out as a promising candidate. It offers a high light yield (58,700 photons/MeV) \cite{Xe-yield}, a primary scintillation characterized by two fast decay constants ($\tau_1$ = 2.2 ns, $\tau_2$ = 27 ns) \cite{Xe-decay} while the decay time associated with recombination is $\tau_3$ = 40 ns \cite{Recomb}, and an excellent scalability due to its liquid state. Consequently, LXe is inherently a brighter and slightly faster scintillator than LYSO. However, its attenuation length is nearly three times longer than LYSO's, and the fraction of photons interacting via the photoelectric effect, which is relevant for PET applications, is a factor of two less, $\sim$22$\%$ \cite{photo_frac_xe, photo_frac_LYSO}, reducing the rate of interactions per unit time, which in turn decreases the overall sensitivity. Additionally, working with LXe presents significant technical challenges, as it requires cooling to -110°C and continuous purification to achieve its highest light yield. This can result in a more complex system, given the safety requirements needed to maintain precise temperature and the pressure control of the LXe.

The idea of constructing a PET scanner based on LXe was first proposed over four decades ago \cite{first-LXe}, and has been explored in several studies since \cite{Chepel,Doke, Solovov}. Recently, the PETALO Collaboration has revisited the concept of a LXe-PET scanner, considering the integration of modern SiPM technology for signal readout. This approach promises better energy resolution and TOF performance compared to previous designs \cite{CTR-PETALO, Cherenkov, MC-intro}.

\section{Experimental setup}

\begin{figure}[htbp]
\centering 
\includegraphics[width=0.60\textwidth,]{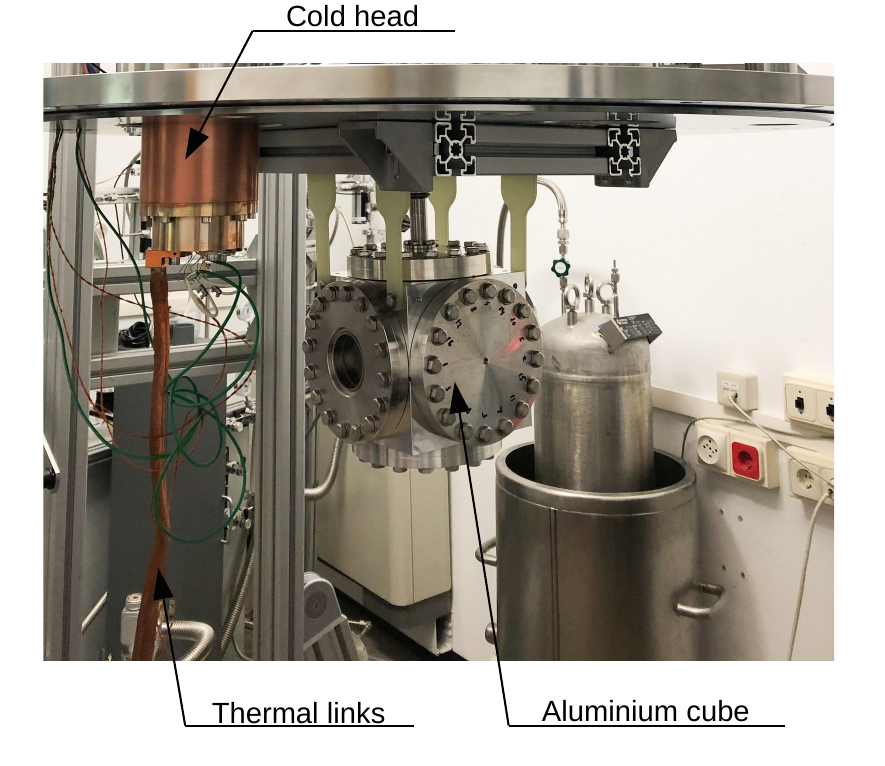}

\caption{\label{fig:PETit} The aluminum cube that keeps the LXe inside when operating and the thermal links attached to the cold head.}
\end{figure}

PETit is the first prototype of PETALO and consists of a metal-aluminum cube (F100-1ALU) from Vacom filled with liquid xenon suspended inside a vacuum vessel. To liquefy the xenon, a Sumitomo CH-110 cold head is coupled to the cube via custom-made copper thermal links, allowing the system to reach xenon's liquefaction temperature of -110ºC. The thermal links are wrapped in polyethylene terephthalate foils to minimize heat dissipation. The vacuum vessel helps maintain the cryogenic temperature of the xenon, by reducing heat transfer. The cube and cooling equipment are shown in Figure \ref{fig:PETit}. Inside the cube, xenon is continuously recirculated in its gas phase using a double diaphragm compressor and purified by passing through a PS4 MT15 R2 hot getter filter from Sigma Technologies. This filter removes impurities such as oxygen, nitrogen and water, which can quench xenon's scintillation light~\cite{xenon-light}.  The gas passes through a heat exchanger when entering and exiting the cube, which precools the gas before liquefaction. 

The prototype is instrumented with two planes of Hamamatsu VUV-sensitive S15779 SiPMs with  a micro-cell pitch of 75 $\mu$m, positioned on opposite sides of the cube and facing each other. Each SiPM has an active area of 5.95 x 5.85 mm$^2$ and they are arranged in arrays of 8 x 8 sensors each, providing a total active surface of 23.04 cm$^2$ on each plane. These SiPMs offer high gain and fast response, which are essential to obtain good time resolution and they also exhibit low dark noise at cryogenic temperatures. They have 30$\%$ photodetection efficiency at the peak wavelength of xenon scintillation and are protected by a quartz window with over $90\%$ transmission at relevant  wavelengths.  

A $^{22}$Na calibration source, emitting pairs of nearly back-to-back gammas with an energy of 511 keV, is placed in the middle of the cube, within a port to avoid contact with the liquid xenon. The radioactive material, with an activity of $\sim$260 kBq at the time of the measurement, consists of a 0.25 mm  diameter sphere, which can be considered point-like, and is encapsulated in a plastic support. 

\begin{figure}[htbp]
\centering 
\includegraphics[width=0.6\textwidth,]{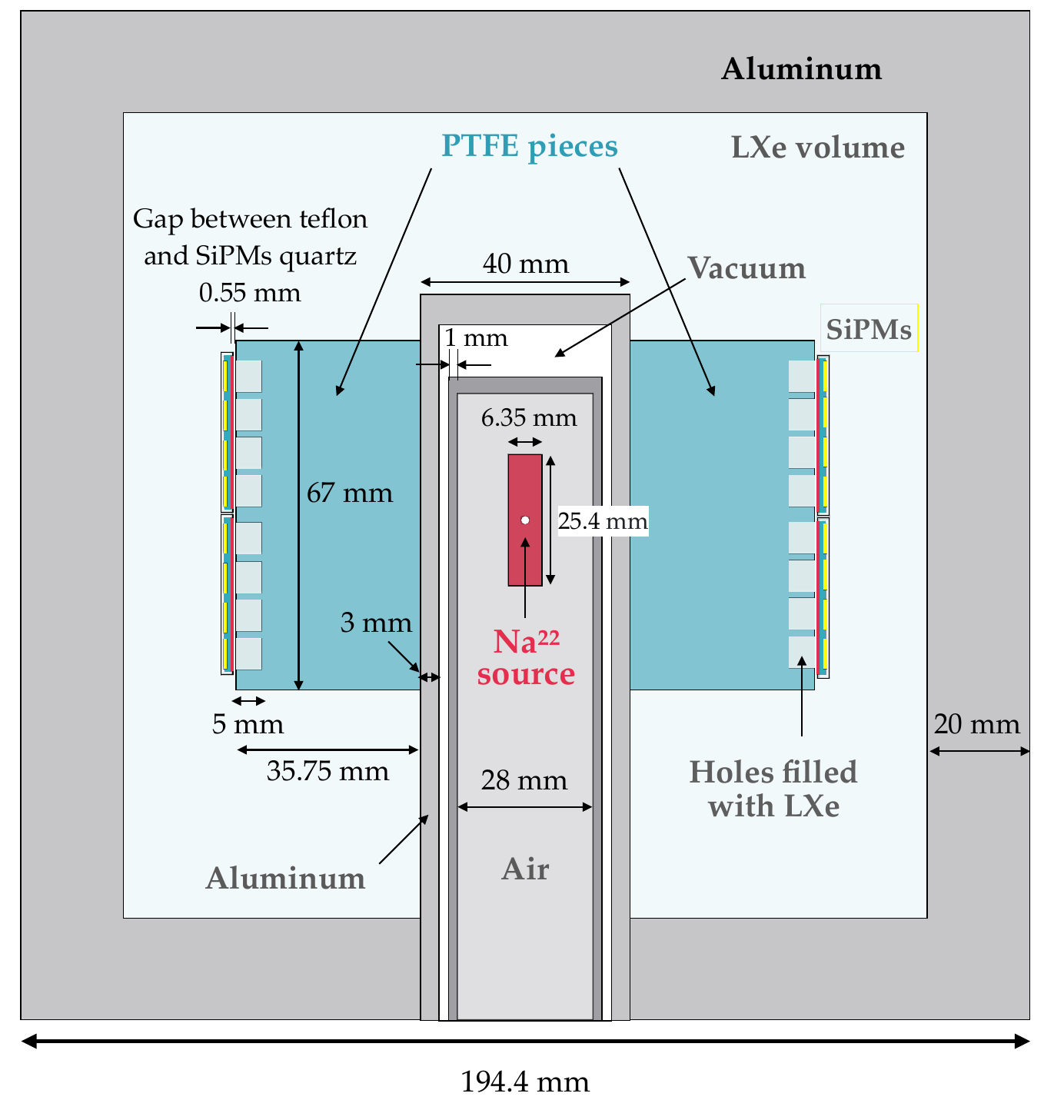}

\caption{\label{fig:sketch} The inner sketch of PETit with the dimensions of the relevant parts.}
\end{figure}

The goal of PETit is to achieve the best possible time resolution, which requires maximizing light collection and minimizing distortions introduced by the depth of interaction. To this end, a PTFE block, highly reflective to VUV wavelengths ($\sim$98$\%$)~\cite{teflon-1, teflon-2}, is placed in front of each SiPM array. The block is machined to form an array of contiguous cells to be filled with liquid xenon, each aligned with a SiPM and having the same transversal dimensions as the SiPM but a depth of 5 mm (see figure \ref{fig:sketch}). The wall thickness between the cells is of only 1.75 mm. At this thickness, optical crosstalk between adjacent cells is expected to be negligible, since measurements show that the VUV light transmittance through 1.75 mm of PTFE is only 0.67$\%$ \cite{teflon-transmitance}, which provides effective optical isolation between neighboring volumes.

The PTFE pieces are mounted to leave a 0.5 mm gap between the SiPM face and the end of each cell. This gap allows liquid xenon to fill the cells completely, ensuring that when 511 keV gamma rays interact with the xenon volume, most of the scintillation light reaches the SiPM aligned with the hole.

The scintillation light detected by the SiPMs is digitized with two TOFPET 2C ASICs from PETsys~\cite{ASIC-1, ASIC-2}, one for each readout plane. These ASICs are mounted on the vacuum side of the cube walls corresponding to the SiPM arrays. 

\subsection{TOFPET 2C ASIC: Time measurement}\label{subsec:PETSYS}

TOFPET2 is a low-power, low-noise ASIC designed for Time-Of-Flight applications, featuring 64 independent channels that provide both charge and time information. To minimize dead time, each channel includes 4 independent analog buffers to store the charge before digitization. A round-robin mechanism selects the buffer for each incoming event. 

The ASIC operates using a dual-threshold trigger with fast dark count rejection~\cite{PETSYS}. The lower threshold measures the event timestamp in the signal's rise edge, while the higher threshold rejects dark counts and integrates the charge to provide the energy measurement. Only the events that cross both thresholds are accepted and digitized; otherwise, they are rejected without introducing any logic dead time. Importantly, both thresholds can be configured independently for each of the 64 channels. Furthermore, each buffer is calibrated separately for energy and time. 

\begin{figure}[htbp]
\centering 
\includegraphics[width=0.85\textwidth,]{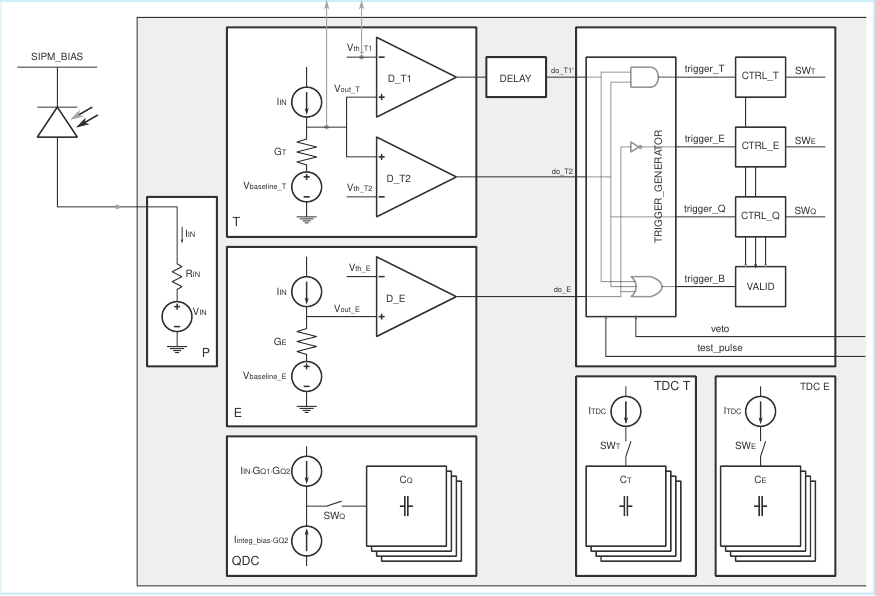}

\caption{\label{fig:TOFPET} Simplified equivalent TOFPET 2C channel from \cite{PETSYS}.}
\end{figure}

The architecture of the TOFPET 2C ASIC ensures that time and charge are processed through independent analog paths. The incoming signal from the SiPM is split into three distinct branches: T for timing, E for energy validation and Q for charge integration (see Figure \ref{fig:TOFPET}). Each branch has its own preamplifier, buffer, and control logic, and digitization is performed through separate analog buffers. 

\begin{figure}[htbp]
\centering 
\includegraphics[width=0.69\textwidth,]{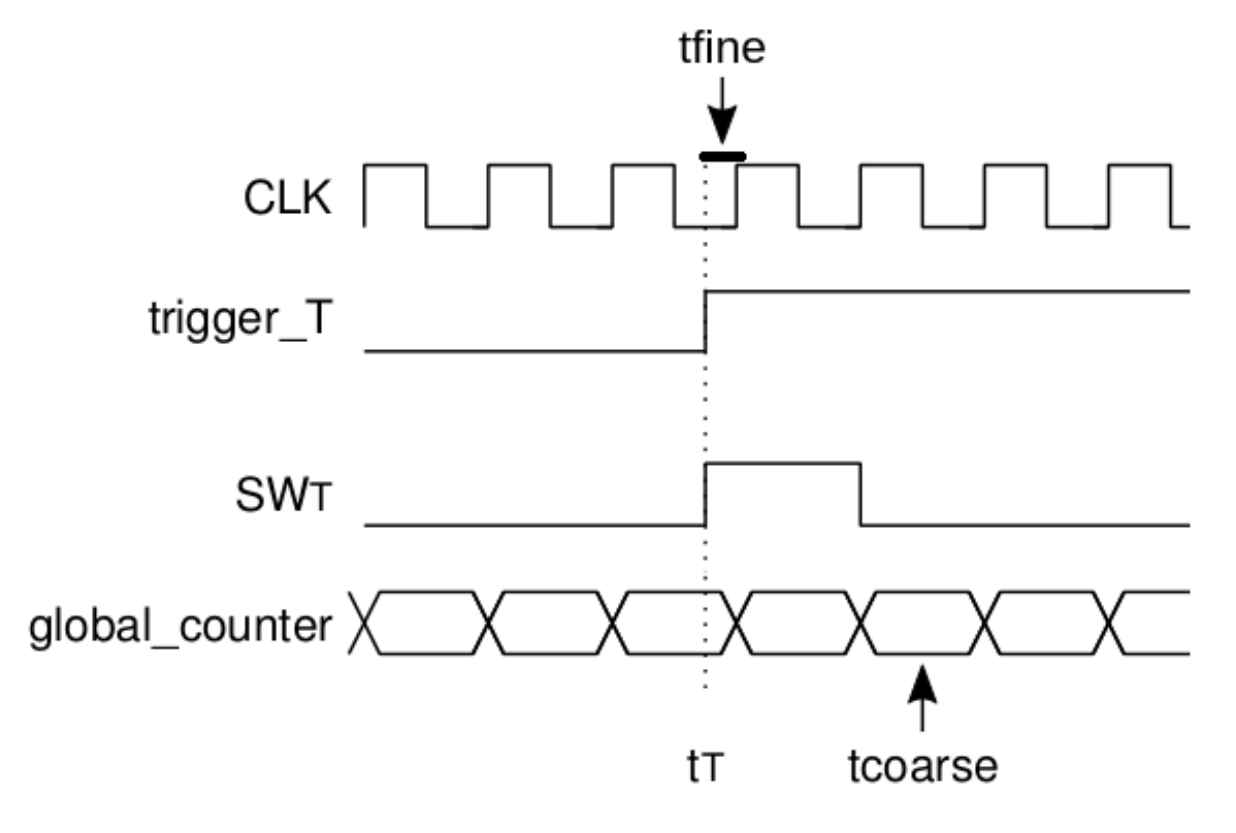}

\caption{\label{fig:timestamp} TDC operation. Reproduced from Reference \cite{PETSYS}.}
\end{figure}

The timestamp is measured using a time-to-digital converter (TDC) integrated in each channel of the ASIC. When the signal crosses the lower threshold, a discriminator generates a digital trigger (trigger$\_$T) that initiates the time measurement process. This process combines a coarse timestamp (tcoarse), obtained from a global counter synchronized with the system clock, and a fine timestamp (tfine) calculated by integrating a constant current into an analog capacitor during a fixed time interval (see figure \ref{fig:timestamp}). Specifically, the trigger closes a switch that allows a defined current to charge the capacitor, and the voltage reached is then digitized to estimate the fine component of the arrival time. This architecture enables time binning down to 30 ps and supports accurate timestamping even at high event rates. The final timestamp can be obtained using equation \ref{eq:timestamp}.

\newpage

\begin{equation}
    \centering
    \text{tT} = \text{tcoarse} - \text{tfine} \cdot TDC_\text{LSB} \quad \text{with} \quad TDC_\text{LSB} = 5 \text{ns}
    \label{eq:timestamp}
\end{equation}

\section{Experimental methods}

During a run, when the charge recorded in a specific channel and buffer crosses both thresholds, the timestamp and the integrated charge of the channel are saved to a file. The data is grouped event by event based on the channel timestamp, and an initial filter is applied to retain only coincidences, i.e., events where at least one sensor from each detection plane is present. Due to the cell structure of the PTFE block, most of the charge is detected by the sensor coupled to the xenon volume where the interaction occurred. Therefore, we single out the sensor with the highest detected charge for each plane, which consistently corresponds to the earliest detection of scintillation light, further confirming its association with the primary interaction site.

\begin{figure}[htbp]
\centering 
\includegraphics[width=0.69\textwidth,]{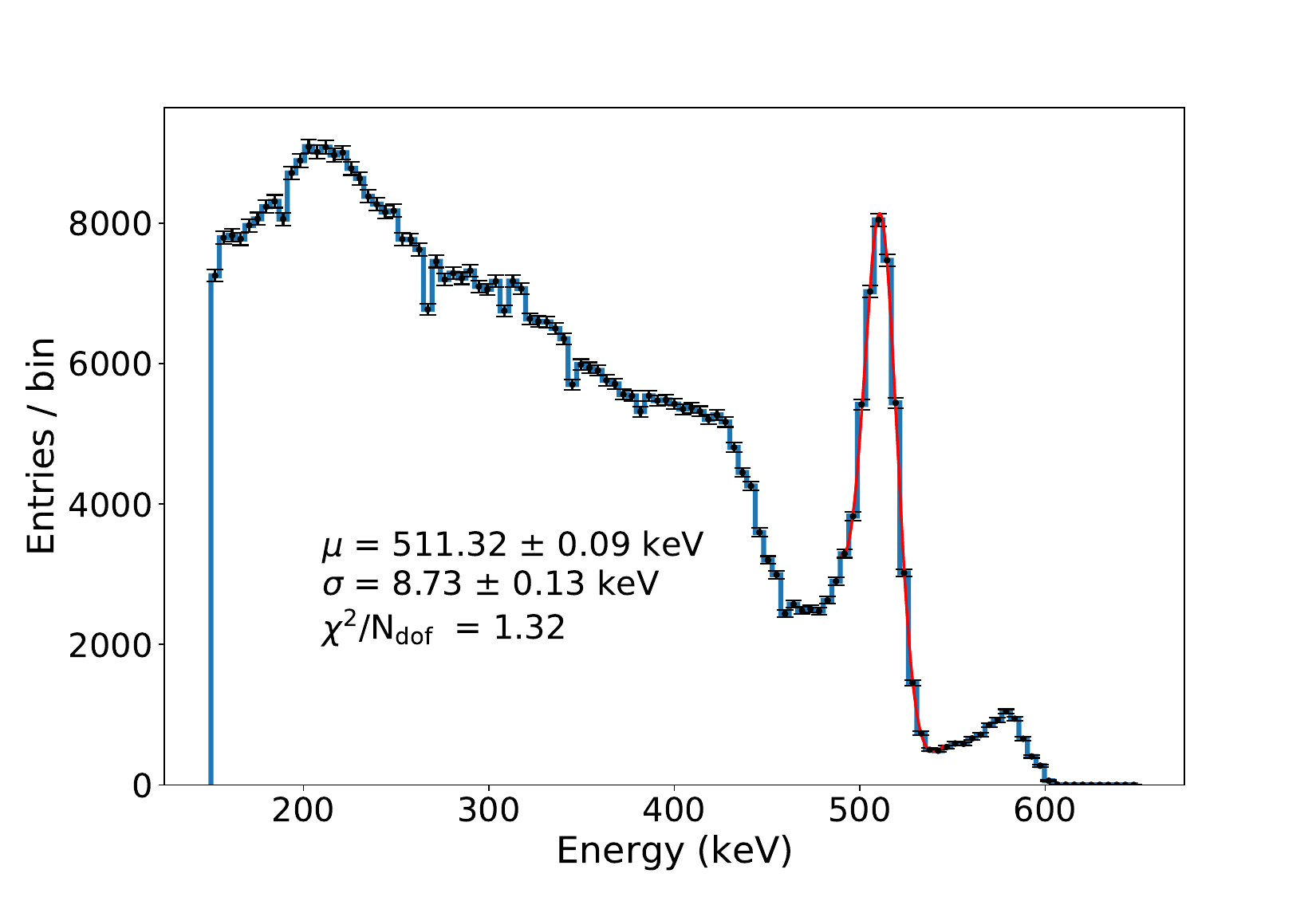}

\caption{\label{fig:energy} Energy spectrum obtained by one of the SiPMs, fitted to a Gaussian function in the photo-peak.}
\end{figure}

Then, we fit the photopeak in the charge spectrum (see Figure \ref{fig:energy}) and a second filter is applied to retain only the events within this peak. We select the events where the energy in both planes falls within the range of the fitted $\mu$ $\pm$ $2\sigma$.

As shown in Figure \ref{fig:energy}, the energy spectrum exhibits a noticeable compression effect at high energies, indicative of saturation in the charge readout. In particular, the position of the Compton edge is shifted to approximately 420 keV, whereas for 511 keV gamma rays it is expected to lie around 340–360 keV. This shift suggests that the integrated charge signal is being clipped due to SiPM saturation under high photon flux conditions.

As discussed previously in subsection  II.A., the TOFPET2C ASIC processes time and charge through separate analog paths. In this architecture, the charge measurement is performed independently from the timing logic, which is based on the leading-edge crossing in the T-branch. Therefore, although saturation affects the charge linearity and distorts the energy spectrum, it does not compromise the accuracy of the timestamp associated with each event, a crucial feature that preserves the validity of the TOF information.

\begin{figure}[htbp]
\centering 
\includegraphics[width=0.69\textwidth,]{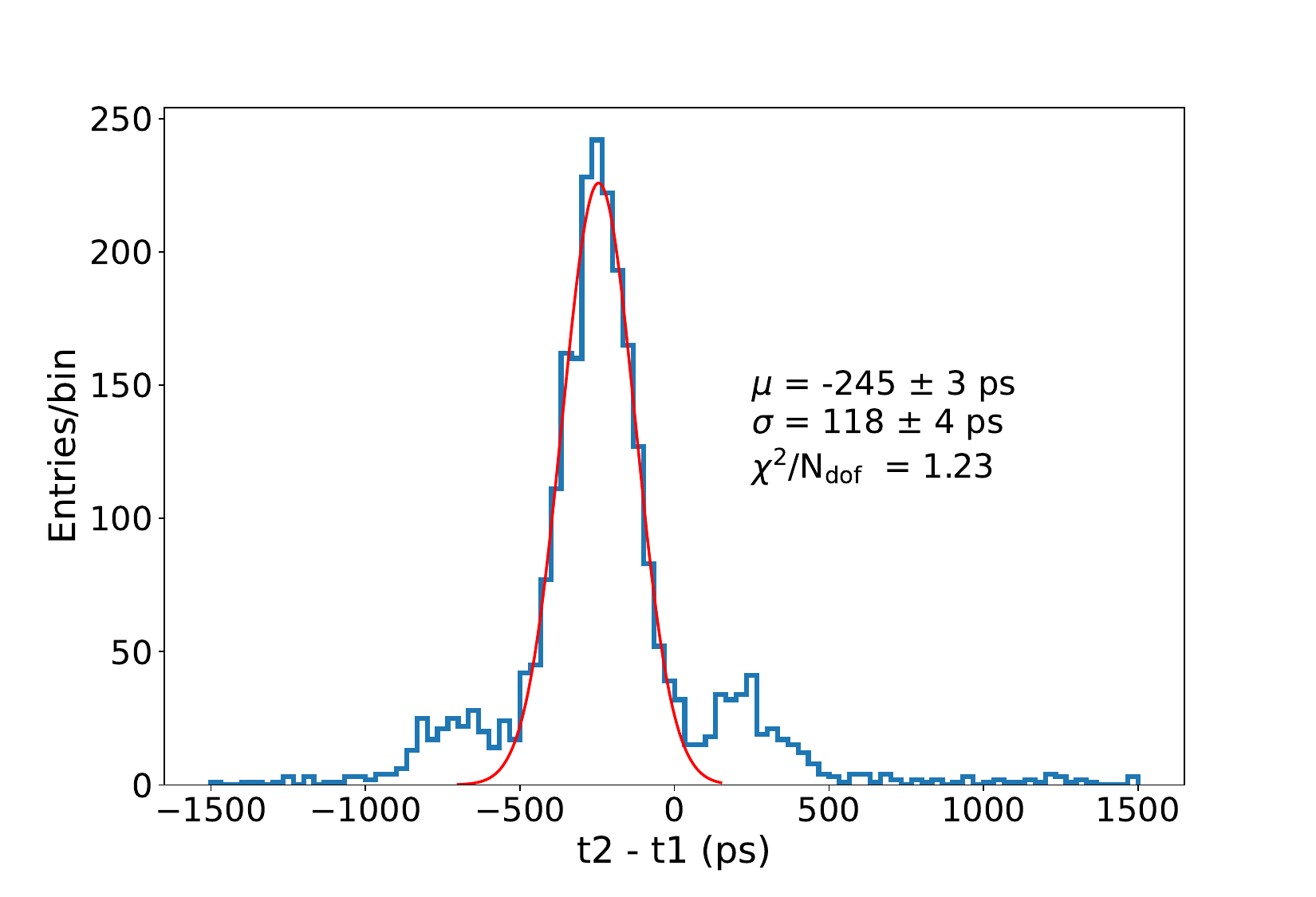}

\caption{\label{fig:CTR_pair_all_tacs} Time difference distribution obtained for a pair of channels.}
\end{figure}

Finally, a third filter is applied to select only the diametrically opposite pairs of channels relative to the calibration source. This step helps reject events where one or both gammas interact with the detector materials before entering the xenon, which would introduce noise in the Time-Of-Flight calculation, since their paths would be deviated. An example of the time difference distribution obtained for a pair of channels is shown in Figure \ref{fig:CTR_pair_all_tacs}.

Beyond the main central peak, the distribution exhibits two smaller peaks approximately 400 ps away from the main one. This feature appears consistently across all pairs of channels, and the secondary peaks are found to be at roughly the same distance. These side peaks are already reported in literature \cite{side-peaks}.

\begin{figure}[h]
\centering 
\includegraphics[width=0.90\textwidth]{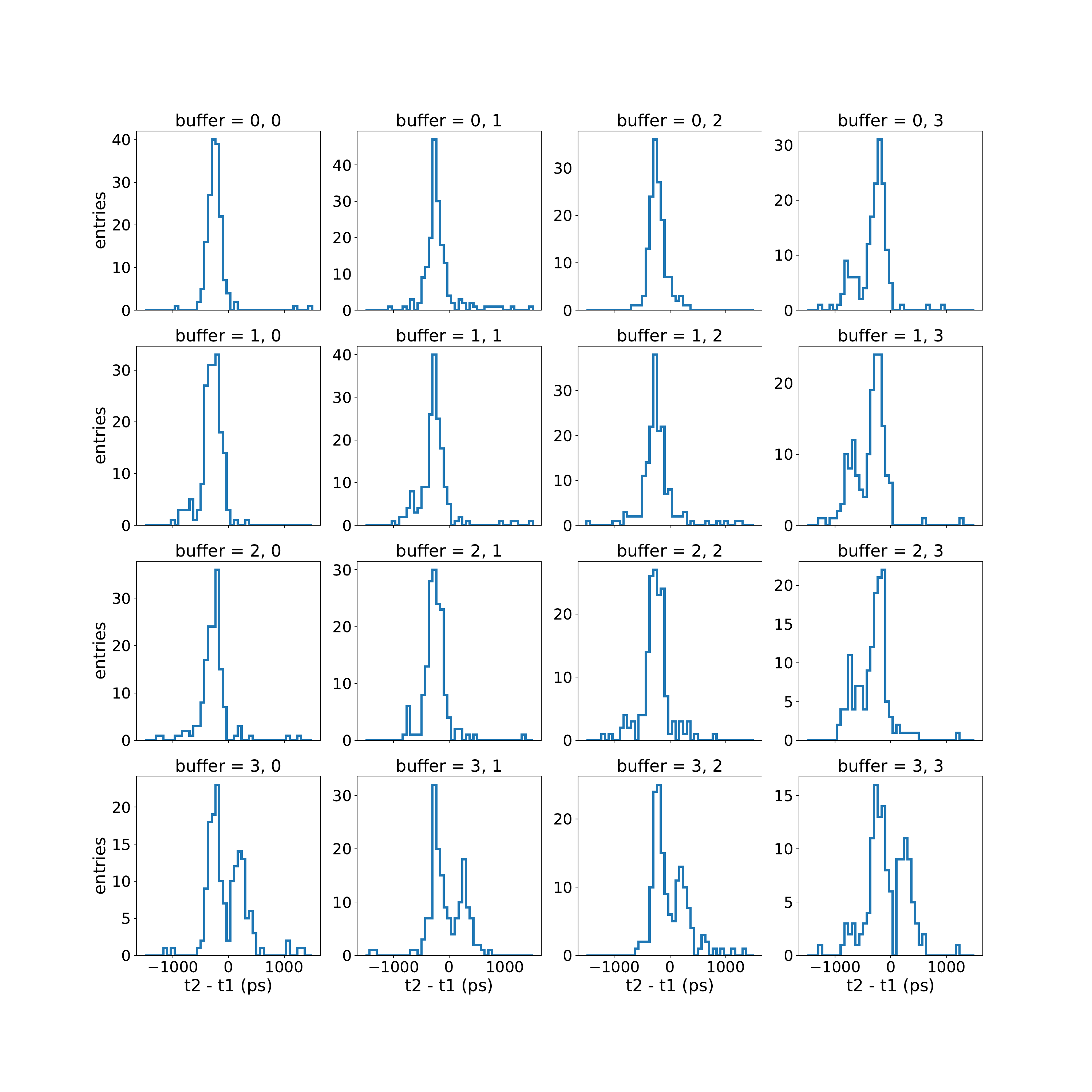}
\caption{\label{fig:CTR_all_tacs_comb} Time difference distributions for a pair of channels for all the buffer combinations.}
\end{figure}

In order to understand the appearance of these outer peaks, we studied the time difference distribution for each possible buffer combination (see Figure \ref{fig:CTR_all_tacs_comb}, which shows a particular pair of channels) finding a relationship between these peaks and the buffer used. When buffer 0 is selected in both channels, the spurious peaks on the sides of the main peak disappear. However, for the other combinations the secondary peaks remain, with the combination involving buffer 3 in both channels showing the highest frequency. 

\begin{figure}[htbp]
\centering 
\includegraphics[width=0.69\textwidth,]{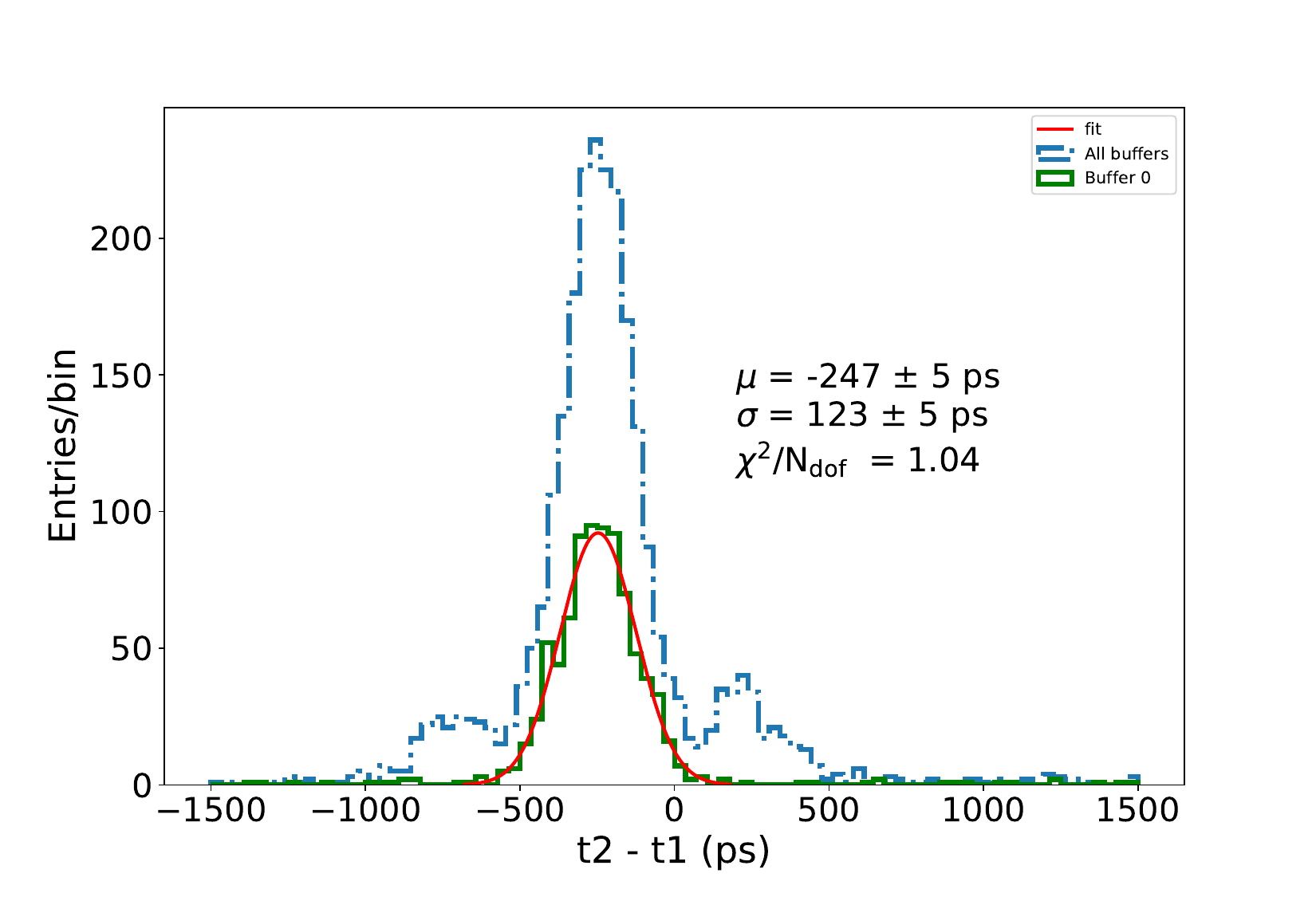}
\caption{\label{fig:CTR_tac_00} Time difference distribution obtained for a pair of channels with buffer 0 compared with the time difference distribution obtained for all buffers. Fit for buffer 0  is shown.}
\end{figure}

This behaviour is likely due to an unresolved issue in the time calibration of the different buffers. Consequently, we decided to filter the data retaining only the events where buffer 0 is used in both channels. The distribution of a pair of channels after applying this last filter is shown in Figure \ref{fig:CTR_tac_00}, fitted to a single Gaussian function.

\begin{figure}[htbp]
\centering 
\includegraphics[width=0.49\textwidth,]{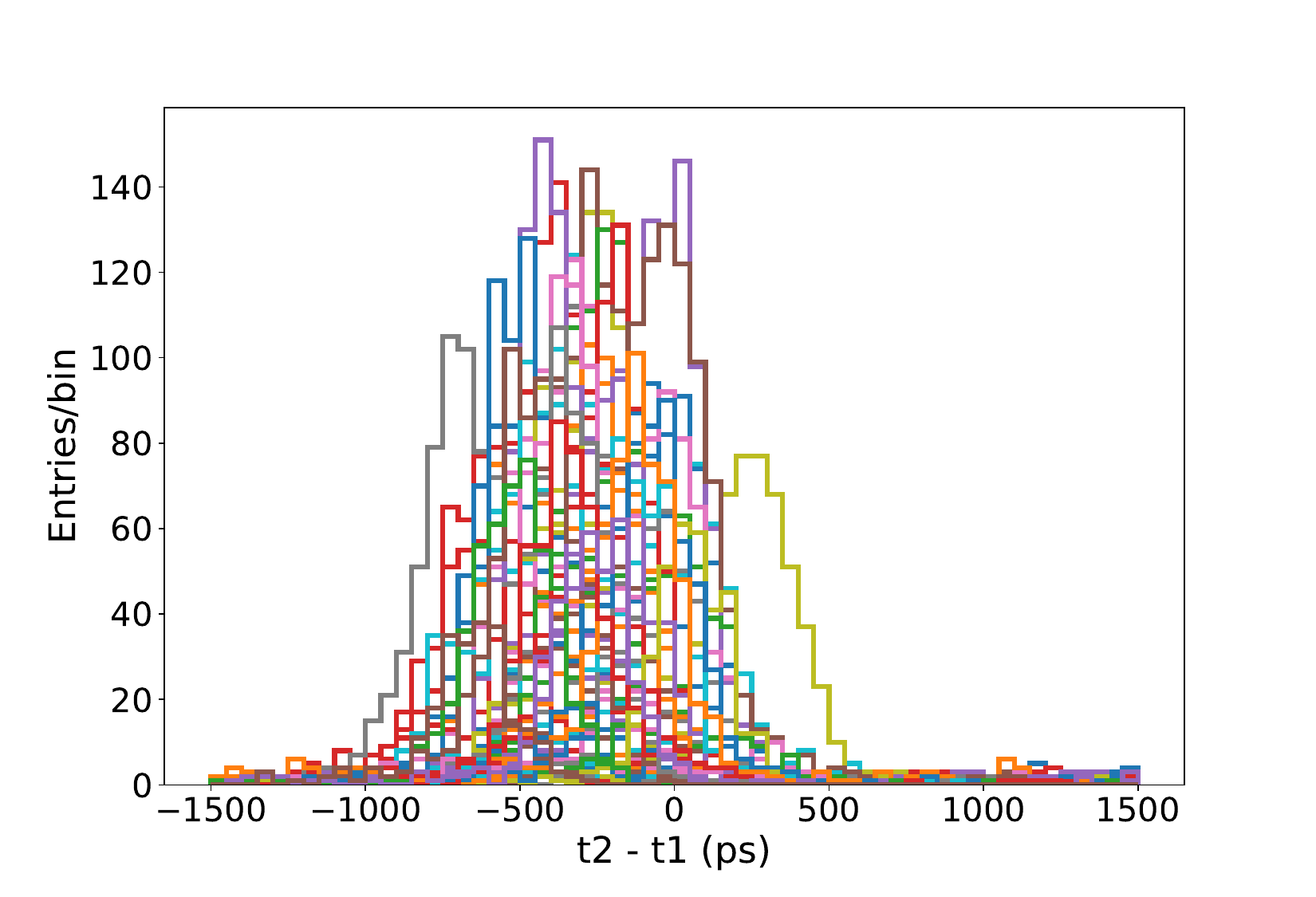}
\includegraphics[width=0.49\textwidth,]{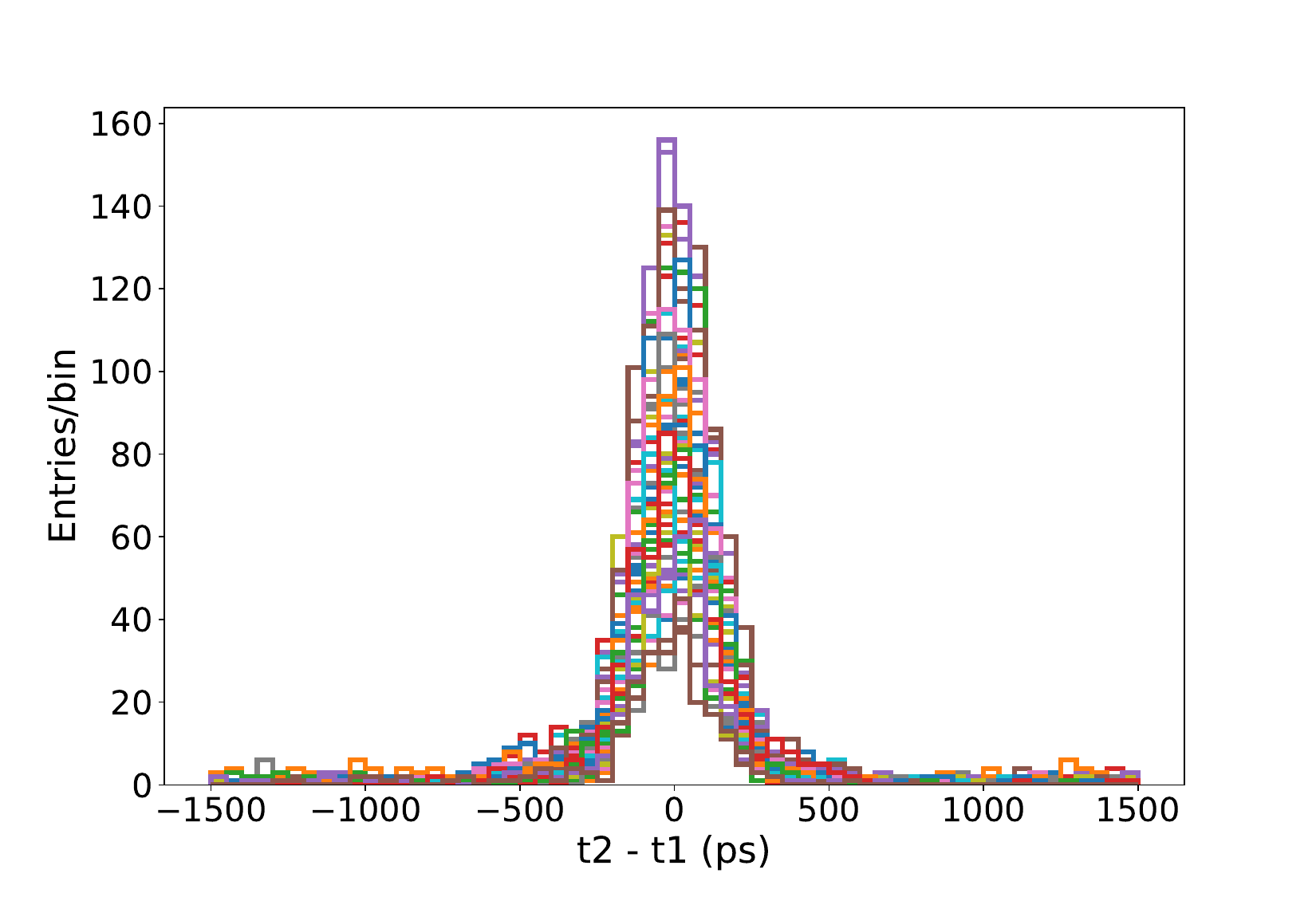}
\caption{\label{fig:CTR_all_channels} Time difference distributions for every pair of channels, selecting only buffer 0 in both channels. Left: raw data. Right: after skew correction.}
\end{figure}

Every pair of channels exhibits a displacement of the mean of the distribution from zero, due to time skew. Time skew in PET scanners refers to the discrepancy in the timing of signal detection between paired detectors that are meant to simultaneously detect gamma photon pairs emitted from a positron annihilation event. Ideally, these photons would be detected at the same time by the detectors on opposite sides of the scanner, therefore the time difference would be centered around zero. However, skews in timing can arise due to several factors, including internal variations in the electronic processing across ASIC channels, differences in the trace routing on the SiPM boards, and possibly small residual mechanical misalignments. We observe variations in the Gaussian centroids as large as 950 ps, as illustrated in Figure \ref{fig:CTR_all_channels}\textit{-left}. This variation cannot be attributed only to routing differences, as we have verified that these are below 1 cm in the worst case, corresponding to a time difference of approximately 70 ps. All evidence points to an ASIC internal deviation in the channel delay calibration as the dominant source. To correct for this effect, we first obtained the Gaussian centroid for each pair of channels and subtracted this value from the time difference. In this way, we used the Gaussian centroids as an offset calibration for the recorded timestamps. Applying the correction, we obtained the histogram shown in Figure \ref{fig:CTR_all_channels}\textit{-right}, which is now centered at 0 ps. 

\newpage

\section{Results and discussion}

\begin{figure}[htbp]
\centering 
\includegraphics[width=0.69\textwidth,]{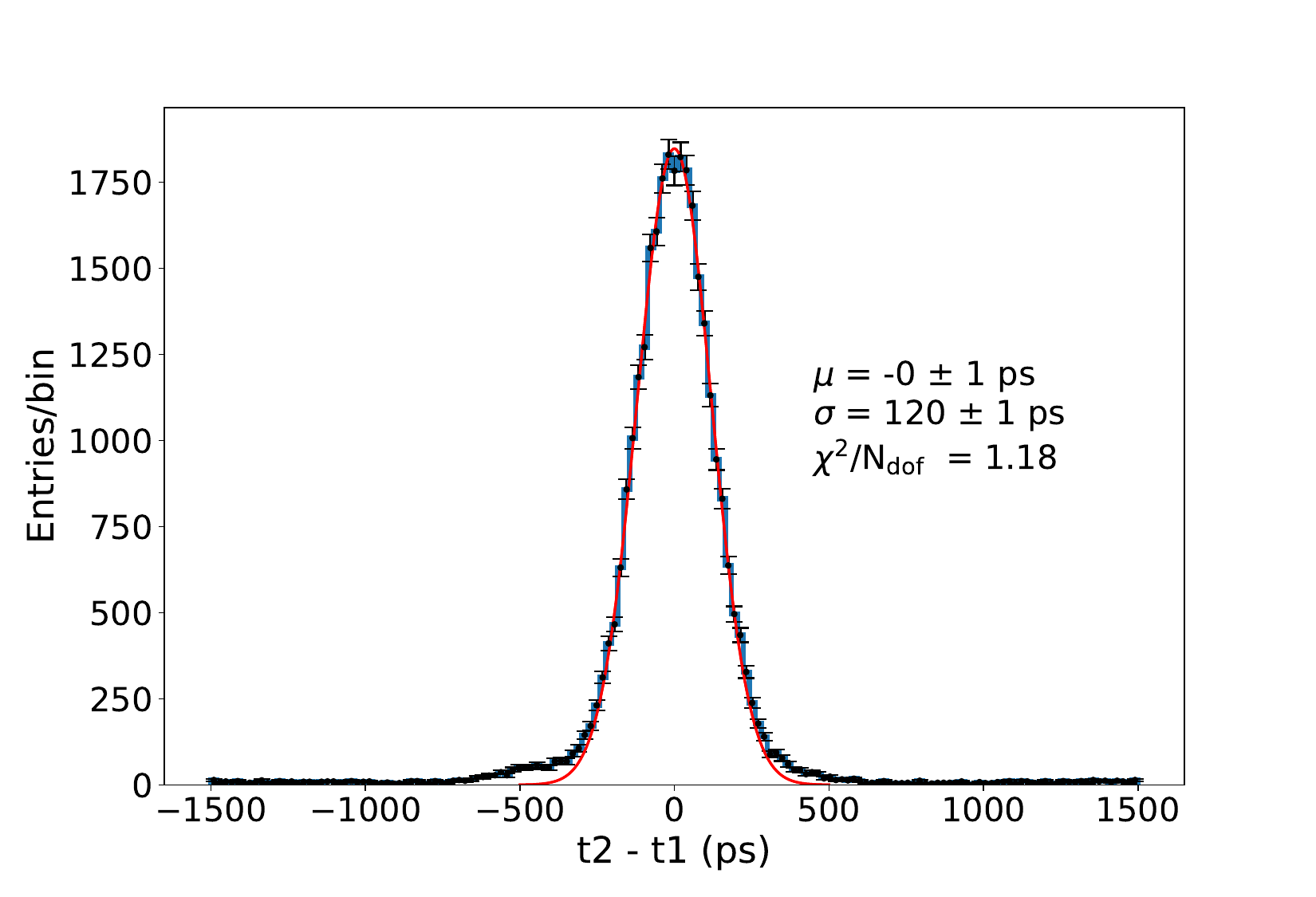}
\caption{\label{fig:CTR_joined} Fit to a Gaussian function on the distribution of the time difference of all channels after time skew correction.}
\end{figure}

After applying the time skew correction to all pairs of channels, we combined the data into a single histogram, as shown in Figure \ref{fig:CTR_joined}. This combined histogram represents the time difference distribution across all channels. By performing a Gaussian fit to this aggregated distribution, we obtained a CTR of 281 $\pm$ 2 ps FWHM. 

As shown in Figure \ref{fig:CTR_joined}, the distribution exhibits non-Gaussian tails, which can likely be attributed to residual contributions from nearby peaks observed in Figure \ref{fig:CTR_tac_00}. Despite applying a buffer 0 filter to suppress these features, some channels may still contribute slightly to them. When summing over all channels, these small residuals can give rise to the observed tails. Moreover, as illustrated in Figure \ref{fig:CTR_all_tacs_comb}, these contributions are not necessarily symmetric, which could explain the larger tail on the lower-value side.

The result obtained is a compelling proof-of-concept of the potential of liquid xenon combined with silicon photomultipliers in TOF-PET technology. The achieved time resolution is competitive with current state-of-the-art PET scanners, such as the Philips Vereos, which features a CTR of 310 ps FWHM \cite{vereos-ctr}. This represents a significant milestone in the development of LXe-based PET technology, highlighting its feasibility as a viable alternative to conventional scintillators like LYSO. The next step will be to measure the CTR with a longer depth of interaction, as is typical in clinical PET scanners, to further validate and optimize this approach for practical applications. 

This result also demonstrates a significant advancement compared to earlier efforts using LXe for TOF-PET, such as Reference \cite{Doke} with a CTR of 552 ps or Reference \cite{Solovov} with a CTR of 0.9 ns, where the use of traditional photodetectors limited the achievable time resolution. The integration of SiPMs in our setup allows us to reach a performance that is now competitive with commercial systems.

Regarding the scalability of the concept to a whole-body PET scanner, several possible detector geometries are under study. For example, a design with a LXe thickness of 6 cm (corresponding to 2 radiation lengths, 2$X_0$) would require approximately 427 kg of xenon. We expect timing resolution to be relatively robust against increases in interaction depth \cite{crys-depth, CTR-DOI}. These ongoing investigations will help assess the performance and feasibility of LXe-based detectors in realistic clinical configurations.

\begin{acknowledgments}
This work was supported by the European Research Council under Grant ID 757829 and is part of the PRE2021-097277 grant, funded by MCIN/AEI/10.13039/501100011033 and the ESF+.

\end{acknowledgments}


\bibliography{apssamp}

\begin{thebibliography}{39}
\providecommand{\natexlab}[1]{#1}
\providecommand{\url}[1]{\texttt{#1}}
\expandafter\ifx\csname urlstyle\endcsname\relax
  \providecommand{\doi}[1]{doi: #1}\else
  \providecommand{\doi}{doi: \begingroup \urlstyle{rm}\Url}\fi

\bibitem[Tomitani(1981)]{image-quality}
T.~Tomitani.
\newblock Image reconstruction and noise evaluation in photon time-of-flight assisted positron emission tomography.
\newblock \emph{IEEE Trans. Nucl. Sci.}, 28:\penalty0 4581--4589, 1981.

\bibitem[et~al.(1980)]{Mullani-1980}
N.~A.~Mullani et~al.
\newblock Feasability of time-of-flight reconstruction in positron emission tomography.
\newblock \emph{J. Nucl. Med.}, 21\penalty0 (11):\penalty0 1095--1097, 1980.

\bibitem[et~al.(1982{\natexlab{a}})]{Super-PETT}
M.~M. Ter-Pogossian et~al.
\newblock {Super PETT I: A Positron Emission Tomograph Utilizing Photon Time-of-Flight Information}.
\newblock \emph{IEEE Trans. Med. Imaging}, 1\penalty0 (3):\penalty0 179--187, 1982{\natexlab{a}}.

\bibitem[et~al.(1982{\natexlab{b}})]{LETI}
R.~Gariod et~al.
\newblock {The LETI positron tomograph architecture and time of flight improvements}.
\newblock In \emph{Proc. of The Workshop on Time of Flight Tomography}, pages 25--29, 1982{\natexlab{b}}.

\bibitem[et~al.(1986)]{Texas}
W.~H.~Wong et~al.
\newblock {Performance characteristics of the University of Texas TOFPET-I PET camera}.
\newblock \emph{J. Nucl. Med.}, 25:05, 1986.

\bibitem[et~al.(1988)]{SP3}
T.~K.~Lewellen et~al.
\newblock {Performance measurements of the SP3000/UW time-of-flight positron emission tomograph}.
\newblock \emph{IEEE Transactions on Nuclear Science}, 35\penalty0 (1):\penalty0 665--669, 1988.

\bibitem[et~al.(1990{\natexlab{a}})]{TTV03}
B.~Mazoyer et~al.
\newblock {Physical characteristics of TTV03, a new high spatial resolution time-of-flight positron tomograph}.
\newblock \emph{IEEE Transactions on Nuclear Science}, 37\penalty0 (2):\penalty0 778--782, 1990{\natexlab{a}}.

\bibitem[et~al.(1990{\natexlab{b}})]{CTR}
K.~Ishii et~al.
\newblock High resolution time‐of‐flight positron emission tomograph.
\newblock \emph{Review of Scientific Instruments}, 61\penalty0 (12):\penalty0 3755--3762, 1990{\natexlab{b}}.

\bibitem[Lewellen(1998)]{CTR-2}
T.~K. Lewellen.
\newblock {Time-of-flight PET}.
\newblock \emph{Seminars in Nuclear Medicine}, 28\penalty0 (3):\penalty0 268--275, 1998.

\bibitem[LYS()]{LYSO}
{LYSO(Ce) properties}.
\newblock Available: https://www.advatech-uk.co.uk/lyso\_ce.html.

\bibitem[LSO()]{LSO}
{LSO(Ce) properties}.
\newblock Available: https://www.advatech-uk.co.uk/lso\_ce.html.

\bibitem[et~al.(2007)]{CTR-LYSO}
S.~Surti et~al.
\newblock {Performance of Philips Gemini TF PET/CT scanner with special consideration for its time-of-flight imaging capabilities}.
\newblock \emph{Society of Nuclear Medicine}, 48\penalty0 (3):\penalty0 471--480, 2007.

\bibitem[et~al.(2011)]{CTR-LSO}
B.~W.~Jakoby et~al.
\newblock {Physical and clinical performance of the mCT time-of-flight PET/CT scanner}.
\newblock \emph{Phys. Med. Biol.}, 56\penalty0 (8):\penalty0 2375, 2011.

\bibitem[et~al.(2019{\natexlab{a}})]{explorer}
R.~Badawi et~al.
\newblock {First Human Imaging Studies with the EXPLORER Total-Body PET Scanner}.
\newblock \emph{J Nucl Med.}, 60\penalty0 (3):\penalty0 299--303, 2019{\natexlab{a}}.

\bibitem[van Sluis~et al.(2019)]{CTR-lowest}
J.~van Sluis~et al.
\newblock {Performance characteristics of the digital biograph vision PET/CT system}.
\newblock \emph{J Nucl Med.}, 60:\penalty0 1031--1036, 2019.

\bibitem[et~al.(2019{\natexlab{b}})]{CTR-5ring}
T.~Pan et~al.
\newblock {Performance evaluation of the 5-ring GE discovery MI PET/CT system using the national electrical manufacturers association NU 2-2012 standard}.
\newblock \emph{Med. Phys.}, 46\penalty0 (7):\penalty0 3025--3033, 2019{\natexlab{b}}.

\bibitem[Chepel and Araujo(2013)]{Xe-yield}
V.~Chepel and H.~Araujo.
\newblock Liquid noble gas detectors for low energy particle physics.
\newblock \emph{JINST}, 8\penalty0 (04):\penalty0 R04001, 2013.

\bibitem[Kubota et~al.(1979)Kubota, Hishida, Suzuki, and Ruan]{Xe-decay}
S.~Kubota, M.~Hishida, M.~Suzuki, and J.~Ruan.
\newblock Dynamical behaviour of free electrons in the recombination process in liquid argon, krypton and xenon.
\newblock \emph{Phys. Rev. B}, 20\penalty0 (8):\penalty0 3486--3496, 1979.

\bibitem[et~al.(2018{\natexlab{a}})]{Recomb}
E.~Hogenbirk et~al.
\newblock Field dependence of electronic recoil signals in a dual-phase liquid xenon time projection chamber.
\newblock \emph{Journal of Instrumentation}, 13\penalty0 (10):\penalty0 P10031--P10031, 2018{\natexlab{a}}.

\bibitem[Chepel et~al.(2002)Chepel, Lopes, Solovov, Marques, and Policarpo]{photo_frac_xe}
V.~Chepel, M.~Lopes, V.~Solovov, R.~Marques, and A.J.P.L Policarpo.
\newblock {Development of liquid xenon detectors for medical imaging}.
\newblock \emph{Technique and Application of Xenon Detectors}, pages 28--40, 2002.

\bibitem[Phunpueok et~al.(2014)Phunpueok, Chewpraditkul, Thongpool, and Aphairaj]{photo_frac_LYSO}
A.~Phunpueok, W.~Chewpraditkul, V.~Thongpool, and D.~Aphairaj.
\newblock {Comparison of Photofraction for LuYAP:Ce, LYSO:Ce and BGO Crystals in Gamma Ray Detection}.
\newblock \emph{Proc. of The 15th Internacional Conference of Internacional Academy of Physical Science}, 2014.

\bibitem[Lavoie(1976)]{first-LXe}
L.~Lavoie.
\newblock Liquid xenon scintillators for imaging of positron emitters.
\newblock \emph{Med. Phys.}, 3\penalty0 (5):\penalty0 283--293, 1976.

\bibitem[et~al.(1999)]{Chepel}
V.~Chepel et~al.
\newblock {The liquid xenon detector for PET: recent results}.
\newblock \emph{IEEE Transactions on Nuclear Science}, 46\penalty0 (4):\penalty0 1038--1044, 1999.

\bibitem[Doke et~al.(2006)Doke, Kikuchi, and Nishikido]{Doke}
T.~Doke, J.~Kikuchi, and F.~Nishikido.
\newblock Time-of-flight positron emission tomography using liquid xenon scintillation.
\newblock \emph{Nucl. Instrum. Methods A}, 569\penalty0 (3):\penalty0 863--871, 2006.

\bibitem[Solovov et~al.(2002)Solovov, Hitachi, Chepel, Lopes, Ferreira~Marques, and Policarpo]{Solovov}
V.~N. Solovov, A.~Hitachi, V.~Chepel, M.~I. Lopes, R.~Ferreira~Marques, and A.~J. P.~L. Policarpo.
\newblock {Detection of Scintillation Light of Liquid Xenon with a LAAPD}.
\newblock \emph{Nucl. Instrum. Meth. A}, 488:\penalty0 572, 2002.

\bibitem[Gómez-Cadenas et~al.(2016)Gómez-Cadenas, Benlloch-Rodríguez, Ferrario, Monrabal, Rodríguez, and Toledo]{CTR-PETALO}
J.J. Gómez-Cadenas, J.M. Benlloch-Rodríguez, P.~Ferrario, F.~Monrabal, J.~Rodríguez, and J.F. Toledo.
\newblock {Investigation of the coincidence resolving time performance of a PET scanner based on liquid xenon: a Monte Carlo study}.
\newblock \emph{JINST}, 11\penalty0 (09):\penalty0 P09011, 2016.

\bibitem[Gómez-Cadenas et~al.(2017)Gómez-Cadenas, Benlloch-Rodríguez, and Ferrario]{Cherenkov}
J.J. Gómez-Cadenas, J.M. Benlloch-Rodríguez, and P.~Ferrario.
\newblock {Monte Carlo study of the coincidence resolving time of a liquid xenon PET scanner, using Cherenkov radiation}.
\newblock \emph{JINST}, 12\penalty0 (08):\penalty0 P08023, 2017.

\bibitem[et~al.(2022)]{MC-intro}
J.~Renner et~al.
\newblock {Monte Carlo characterization of PETALO, a full-body liquid xenon-based PET detector}.
\newblock \emph{JINST}, 17\penalty0 (05):\penalty0 P05044, 2022.

\bibitem[et~al.(2005)]{xenon-light}
A.~Baldini et~al.
\newblock Absorption of scintillation light in a 100 l liquid xenon-ray detector and expected detector performance.
\newblock \emph{Nucl. Instrum. Meth. A}, 545\penalty0 (3):\penalty0 753--764, 2005.

\bibitem[et~al.(2013)]{teflon-1}
D.~Akerib et~al.
\newblock {Technical results from the surface run of the LUX dark matter experiment}.
\newblock \emph{Astroparticle Physics}, 45:\penalty0 34--43, 2013.

\bibitem[et~al.(2004)]{teflon-2}
M.~Yamashita et~al.
\newblock {Scintillation response of liquid Xe surrounded by PTFE reflector for gamma rays}.
\newblock \emph{Nuclear Instruments and Methods in Physics Research Section A: Accelerators, Spectrometers, Detectors and Associated Equipment}, 535\penalty0 (3):\penalty0 692--698, 2004.

\bibitem[Cichon et~al.(2020)Cichon, Eurin, Jörg, Undagoitia, and Rupp]{teflon-transmitance}
D.~Cichon, G.~Eurin, F.~Jörg, T.~Marrodán Undagoitia, and N.~Rupp.
\newblock {Transmission of xenon scintillation light through PTFE}.
\newblock \emph{JINST}, 15\penalty0 (09):\penalty0 P09010, 2020.

\bibitem[et~al.(2018{\natexlab{b}})]{ASIC-1}
V.~Herrero-Bosch et~al.
\newblock {PETALO read-out: a novel approach for data acquisition systems in PET applications}.
\newblock \emph{2018 IEEE Nuclear Science Symposium and Medical Imaging Conference (NSS/MIC)}, 2018{\natexlab{b}}.

\bibitem[et~al.(2016)]{ASIC-2}
A.~Di~Francesco et~al.
\newblock {TOFPET2: a high-performance ASIC for time and amplitude measurements of SiPM signals in time-of-ﬂight applications}.
\newblock \emph{JINST}, 11\penalty0 (3):\penalty0 C03042, 2016.

\bibitem[PET(Revision 13)]{PETSYS}
{TOFPET 2C/D SiPM readout ASIC}.
\newblock \emph{PETsys Electronics}, 29, Revision 13.

\bibitem[Nadig et~al.(2022)Nadig, Yusopova, Radermacher, Schug, Weissler, Schulz, and Gundacker]{side-peaks}
V.~Nadig, M.~Yusopova, H.~Radermacher, D.~Schug, B.~Weissler, V.~Schulz, and S.~Gundacker.
\newblock {A Comprehensive Study on the Timing Limits of the TOFPET2 ASIC and on Approaches for Improvements}.
\newblock \emph{IEEE Transactions on Radiation and Plasma Medical Sciences}, 6:\penalty0 1--1, 2022.

\bibitem[ver()]{vereos-ctr}
{Vereos {PET/CT}}.
\newblock Available: https://www.usa.philips.com/healthcare/product/HC882446/vereos-digital-petct-proven-accuracy-inspires-confidence.

\bibitem[Gundacker et~al.(2014)Gundacker, Knapitsch, Auffray, Jarron, Meyer, and Lecoq]{crys-depth}
S.~Gundacker, A.~Knapitsch, E.~Auffray, P.~Jarron, T.~Meyer, and P.~Lecoq.
\newblock Time resolution deterioration with increasing crystal length in a {TOF-PET} system.
\newblock \emph{Nuclear Instruments and Methods in Physics Research Section A: Accelerators, Spectrometers, Detectors and Associated Equipment}, 737:\penalty0 92--100, 2014.

\bibitem[Brown et~al.(2014)Brown, Gundacker, Taylor, Tummeltshammer, Auffray, Lecoq, and Papakonstantinou]{CTR-DOI}
M.~S. Brown, S.~Gundacker, A.~Taylor, C.~Tummeltshammer, E.~Auffray, P.~Lecoq, and I.~Papakonstantinou.
\newblock Influence of depth of interaction upon the performance of scintillator detectors.
\newblock \emph{PloS one}, 9\penalty0 (5):\penalty0 e98177, 2014.

\end{thebibliography}

\end{document}